\documentclass[iop,apjl,numberedappendix]{emulateapj}
\usepackage[latin1]{inputenc}
\usepackage{amssymb}
\usepackage{amsmath}
\usepackage{graphicx}
\usepackage{xspace}
\usepackage{natbib}
\usepackage{url}
\usepackage{paralist}
\usepackage{epsfig}

\def\cm{{\rm\thinspace cm}}

\def\erg{{\rm\thinspace erg}}

\def\keV{{\rm\thinspace keV}}
\def\km{{\rm\thinspace km}}

\def\Mpc{{\rm\thinspace Mpc}}

\def\s{{\rm\thinspace s}}

\def\pcmcu{\hbox{$\cm^{-3}\,$}}

\def\ergcmps{\hbox{$\erg\cm\s^{-1}\,$}}

\def\kmps{\hbox{$\km\s^{-1}\,$}}

\def\pcmsq{\hbox{$\cm^{-2}\,$}}
\def\pcmcu{\hbox{$\cm^{-3}\,$}}

\def\kmpspMpc{\hbox{$\kmps\Mpc^{-1}$}}

\slugcomment{ }

\shorttitle{MCMC investigation of spin in NGC3783}
\shortauthors{C.~S.~Reynolds et al.}

\begin{document}


\title{A Monte Carlo Markov Chain based investigation of black hole spin in the active galaxy NGC~3783}




\author{Christopher~S.~Reynolds\altaffilmark{1,2}, Laura~W.~Brenneman\altaffilmark{3}, Anne~M.~Lohfink\altaffilmark{1,2}, Margaret~L.~Trippe\altaffilmark{1,2}, Jon~M.~Miller\altaffilmark{4}, Andrew~C.~Fabian\altaffilmark{5}, Michael~A.~Nowak\altaffilmark{6}}
\email{chris@astro.umd.edu}
\altaffiltext{1}{Department of Astronomy, University of Maryland, College Park, MD 20742-2421, USA; alohfink@astro.umd.edu}
\altaffiltext{2}{Joint Space-Science Institute (JSI), College Park, MD 20742-2421, USA}
\altaffiltext{3}{Harvard-Smithsonian Center for Astrophysics, 60 Garden Street, Cambridge, MA}
\altaffiltext{4}{Department of Astronomy, University of Michigan, Ann Arbor, MI}
\altaffiltext{5}{Institute of Astronomy, Madingley Road, Cambridge, CB3 OHA, UK.}
\altaffiltext{6}{MIT Kavli Institute for Astrophysics, Cambridge, MA 02139, USA}


\begin{abstract}
The analysis of relativistically broadened X-ray spectral features from the inner accretion disk provides a powerful tool for measuring the spin of supermassive black holes (SMBH) in active galactic nuclei (AGN).   However, AGN spectra are often complex and careful analysis employing appropriate and self-consistent models are required if one is to obtain robust results.   In this paper, we revisit the deep July-2009 {\it Suzaku} observation of the Seyfert galaxy NGC~3783 in order to study in a rigorous manner the robustness of the inferred black hole spin parameter.  Using Monte Carlo Markov Chain (MCMC) techniques, we identify a (partial) modeling degeneracy between the iron abundance of the disk and the black hole spin parameter.   We show that the data for NGC~3783 strongly require both supersolar iron abundance ($Z_{\rm Fe}=2-4\,Z_\odot$) and a rapidly spinning black hole ($a>0.88$).  We discuss various astrophysical considerations that can affect the measured abundance.  We note that, while the abundance enhancement inferred in NGC~3783 is modest, the X-ray analysis of some other objects has found extreme iron abundances.   We introduce the hypothesis that the radiative levitation of iron ions in the innermost regions of radiation-dominated AGN disks can enhance the photospheric abundance of iron.  We show that radiative levitation is a plausible mechanism in the very inner regions of high accretion rate AGN disks.  
\end{abstract}



\section{Introduction}\label{intro}

The X-ray spectra of active galactic nuclei (AGN) often display relativistically broadened spectral features from the innermost regions of the black hole accretion disk.  Analysis of these features, most notably the broad iron line, is one of the most powerful tools that we currently possess for studying the region close to the supermassive black hole (SMBH) where the gravitational field is relativistic (Fabian et al. 1995; Reynolds \& Nowak 2003; Miller 2007).   As such, essentially all robust measurements of SMBH spin are currently derived from an analysis of these spectral features (Dabrowski et al. 1997; Brenneman \& Reynolds 2006; Reynolds et al. 2012).  

However, the X-ray spectra of AGN are often complex and careful modeling is required if one is to obtain robust results.  The broad iron line results from fluorescence in the surface layers of the optically-thick accretion disk when it is irradiated by an external source of hard X-rays (the corona or the base of a jet).   When using high signal-to-noise spectra to study black hole spin, it is insufficient to study the broad iron line in isolation; one must self-consistently model the full spectrum of X-rays that are reflected/reprocessed by the accretion disk (George \& Fabian 1991).   Photoionization of the accretion disk surface must be modeled (Ballantyne, Fabian \& Ross 2002; Ross \& Fabian 2005; Ballantyne, McDuffie \& Rusin 2011), and the possibility of non-solar abundances must be considered (Reynolds, Fabian \& Inoue 1995).   While acknowledging our uncertainties about the nature of the primary X-ray source in AGN, the physicality of the irradiation profile $\epsilon(r)$ across the disk must be respected.   For example, extremely centrally concentrated irradiation profiles ($\epsilon\propto r^{-q}, q>5$) are difficult to understand {\it unless} the black hole is rapidly-rotating.  Spectral variability presents yet another complicating factor, often putting us in the position of having to model a time-averaged spectrum that has been collected over a time-frame where we know that spectral variability has occurred.    Finally, of course, attention must be paid to modeling the photoelectric absorption features that sometimes dominate the soft X-ray spectra of AGN and may itself be time-variable.

In this paper, we revisit the case of the bright Seyfert 1.5 galaxy NGC~3783 ($z=0.00973$; Theureau et al. 1998).   Using a deep (250\,ks) {\it Suzaku} observation, Brenneman et al. (2011; hereafter B11) identified strongly broadened spectral features and concluded that the black hole is rapidly-spinning in the prograde sense (with respect to the accretion disk), with a dimensionless spin parameter $a>0.88$ at the 99\% confidence level.    Reis et al. (2012) examined spectral variability during this same observation.  They found that, while the strong warm absorber showed no signs of variability, there were significant intra-observation changes in the 15--40keV/2--10keV color which they modeled as ionization changes within the disk.   However, they showed that the overall conclusions of B11 were unaffected by this variability and, again, a rapidly rotating black hole was required.    However, using the same dataset, Patrick et al. (2011; hereafter P11) obtained a different result, preferring a slowly or retrograde spinning black hole $a<-0.04$.   Clearly, this discrepancy must be addressed and the robustness of the spin in this (and other AGN) assessed.

We identify the treatment of the iron abundance as the key that underlies the discrepancies in the published analyses of NGC~3783; P11 assume that the accretion disk possesses cosmic iron abundance whereas B11 allow for non-solar abundances.   Revisiting these data with a Monte Carlo Markov Chain (MCMC) analysis, we  find that black hole spin and iron abundance are statistically correlated variables.   We show that the data strongly support a super-solar iron abundance ($Z_{\rm Fe}=2.3-4.1Z_\odot$; 90\% confidence) and a rapidly spinning black hole ($a>0.89$; 90\% confidence) in this object.    While the iron abundance found in NGC~3783 is completely in line with the general metallicity enhancements of AGN inferred from optical/UV spectroscopy, we note that the X-ray reflection technique has found extreme iron abundances in several narrow line Seyfert 1 galaxies.   We suggest that radiative levitation can enrich the photosphere of the inner accretion disk with iron, leading to {\it apparent} extreme iron abundances.

This paper is organized as follows.  Section~2 gives a brief description of the {\it Suzaku} data and its reduction.  Section~3 describes the spectral modeling of these data, and the results from this modeling are presented in Section~4.    The implications of our study are discussed Section~5, where we also make a preliminary assessment of the ability of radiative levitation to enhance the photospheric abundances of the innermost accretion disk.   Throughout this paper, luminosities and distances are calculated using a $\Lambda$CDM cosmological model with $H_0=71\kmpspMpc$, $\Omega_\Lambda=0.73$ and $\Omega_M=0.27$ (Komatsu et al. 2011).  For a redshift of $z=0.00973$, this results in a luminosity distance to NGC~3783 of 41.4\,Mpc.     Unless otherwise stated, all errors in this paper are quoted at the 90\% confidence level for one interesting parameter.   

\section{Observation and data reduction}

{\it Suzaku} observed NGC3783 quasi-continuously for the period 2009 July 10--15.  The source was placed at the nominal aimpoint for the Hard X-ray Detector (HXD).   After filtering the data for Earth occultations, South Atlantic Anomaly passages, and other high background events, we obtained a total on-source exposure of 210\,ks spread over 360\,ks of wall-clock time.   The data reduction exactly follows that described in B11 except for the use of updated calibration and background files (as of late-Fall 2011).  We use data from all three operating X-ray Imaging Spectrometers (XIS0/XIS1/XIS3), combining the spectra from the two front illuminated (FI) detectors (XIS0 and XIS3) into a single spectrum.  When fitting spectra, the flux cross-normalization between the XIS-FI spectrum and the back illuminated (XIS1) spectrum is left as a free parameter and is found to be approximately 1.03 (the same as with the older calibrations used by B11).    Hard X-ray coverage is provided by the HXD/PIN instrument, reduced as per the {\it Suzaku} ABC guide using the relevant ``tuned" non X-ray background event file.  During spectral fitting, the PIN/XIS-FI flux cross-normalization was fixed at 1.18 (Suzaku Memo 2008-06\footnote{http://www.astro.isas.jaxa.jp/suzaku/doc/suzakumemo/suzakumemo-2008-06.pdf}).  In B11, we investigate the effects of allowing this cross-normalization to be a free parameter, finding that it adopts a value of $1.15\pm 0.07$ with rather little impact on other aspects of the fit.   

The spectral analyses presented in this paper are performed jointly on the simultaneous XIS-FI, XIS1, and HXD/PIN spectra.  We use XIS data in the 0.7--10\,keV band, excluding the 1.5--2.5\,keV region due to the presence of well known calibration artifacts associated with the X-ray mirror and Si-edge in the detector.   XIS data were rebinned to a mininum of 20 photons per energy bin to facilitate the use of $\chi^2$ statistics, and PIN data were regrouped to a minimum signal-to-noise ratio of 5.  

\section{Spectral modeling}

The spectrum of NGC~3783 is complex and its description requires a multi-component model.   We begin by discussing the basic components and parameters of the base model.  At the end of this section, we delineate the specific variants of the model that we shall employ.

The primary X-ray continuum is described by a power-law with photon index $\Gamma$.   X-ray reflection from distant neutral material (the outer accretion disk or the dust molecular torus) is described using the {\tt pexmon} model (Nandra et al. 2007; B11) that includes the iron K$\alpha$/K$\beta$ fluorescent emission lines and the associated Compton reflection continuum.  The strength of this neutral reflection is characterized by the reflection fraction, ${\cal R}$, normalized such that ${\cal R}=1$ corresponds to a reflector that subtends a solid angle of $2\pi$ as seen from the source.  To model an additional narrow feature in the iron K-band, we need to add a narrow {\sc Fe\,xxvi} emission line at $6.97$\,keV which presumably arises from highly photoionized optically-thin matter close to the AGN.    X-ray reflection from the ionized inner accretion disk is described with the {\tt reflionx} model (Ross \& Fabian 2005) convolved with the {\tt relconv} kernel in order to include relativistic Doppler and gravitational redshifts (Dauser et al. 2010).  The rest-frame reflection spectrum of the disk is characterized by its ionization parameter $\xi$, and the iron abundance relative to cosmic $Z_{\rm Fe,disk}$.   The relativistic smearing of the disk spectrum depends upon the spin of the black hole $a$, the inclination of the inner disk $i$, the radius of the inner edge of the accretion disk (set to the innermost stable circular orbit [ISCO] for that spin), and the irradiation/emissivity profile of the disk (set to be a broken power law, breaking from $r^{-q_1}$ to $r^{-q_2}$ at radius $r=r_{br}$).  Provided that $q_2>2$, the line profile is insensitive to a (sufficiently large) outer radius; we fix $r_{\rm out}=400r_g$, where $r_g\equiv GM/c^2$ is the gravitational radius.

The multi-component spectrum of NGC3783 is strongly affected by photoionized absorption which, based upon {\it Chandra}/HETG data (Kaspi et al. 2002, Netzer et al. 2003, Brenneman et al. 2011), we model using a three-zone warm absorber (with column densities $N_i$ and $\xi_i$, $i=1,2,3$).  We allow a fraction $f$ of the continuum emission to scatter/leak around the photoionized absorbers.  We use the XSTAR code (Kallman \& Bautista 2001) to compute a grid of absorption spectra on the 2-dimensional $(\xi,N)$-plane, sampling the parameter space at 20 logarithmically-spaced points in column density (from $N=10^{20}\pcmsq$ to $N=10^{24}\pcmsq$) and 20 logarithmically-space points in ionization parameter (from $\xi=1\ergcmps$ to $\xi=10^4\ergcmps$).  For the purpose of defining the ionization balance in the absorbing zone, the photoionizing spectrum is fixed at $\Gamma=2$, a good approximation to the actual primary X-ray spectrum of NGC3783.  The velocity dispersion of the absorbing gas is set to $200\kmps$.  Motivated by P11, we experimented with using a higher velocity dispersion (1000\kmps) for the most highly ionized absorber but found that the results were unchanged.

\begin{figure}[t]
\centerline{
\psfig{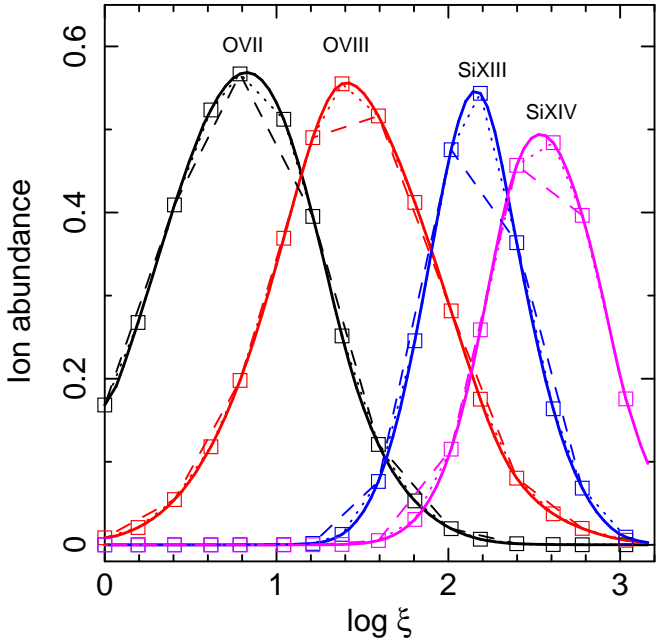}
}
\caption{Illustration of the importance of $\xi$-resolution in constructing warm absorber tables.    The solid lines show the abundance of four important ions ({\sc O\,vii}, {\sc O\,viii}, {\sc Si\,xiii}, and {\sc Si\,xiv}) relative to  the total abundance of that element as a function of the ionization parameter $\xi$; these curves are computing using XSTAR assuming a plasma density of $10^{10}\pcmcu$ and a power-law ionizing spectrum with photon index $\Gamma=2$ extending from 1\,Ry to 1000\,Ry.  The squares connected by dotted lines illustrate the accuracy with which linear interpolation can reproduce these curves with a 5-samples-per-decade criterion ($\Delta\log\xi=0.2$).    The dashed lines illustrate a linear interpolation using just 2.5-samples-per-decade.   The lower resolution sampling leads to substantial errors in the reconstructed ion abundance (and hence opacities).  }
\label{fig:ion_abundance}
\end{figure}

We note that, when employing tabulated models to describe a strong warm absorber, it is crucial to sample the parameter space with sufficient resolution.   Of special importance is the $\xi$-resolution of the tabulated grid.    The abundance of a given ion (say {\sc O\,vii} or {\sc Si\,XIV}) displays a broad peak as a function of $\xi$.   This peak must be resolved by the tabulated grid if one wishes to obtain accurate spectra when interpolating to intermediate values of $\xi$.   This is illustrated in Fig.~\ref{fig:ion_abundance} which shows the ion abundance (relative to the total abundance of that element) for {\sc O\,vii}, {\sc O\,viii}, {\sc Si\,xiii}, and {\sc Si\,xiv} as a function of ionization parameter $\xi$, calculated using XSTAR assuming a plasma density of $10^{10}\pcmcu$ and a power-law ionizing spectrum with photon index $\Gamma=2$ extending from 1\,Ry to 1000\,Ry.  Sampling $\xi$ with a resolution $\Delta\log\xi=0.2$ and linearly extrapolating between sample points (as does XSPEC) produces estimates of the ion abundances accurate to better than 5\%, translating directly into interpolated warm absorber grids that will possess opacities accurate to better than 5\%.      These considerations drive the construction of our absorber grids; we have 5 models per decade of $\xi$ across 4 decades of $\xi$.   It does raise serious concerns about results obtained using the XSPEC model {\it zxipcf}; the grid underlying this model only samples 12 points across 9 decades of $\xi$ and hence is clearly undersampled.    This resolution issue may also be another difference between B11 and P11; the XSTAR tables used for the P11 analysis either sample 5 decades of $\xi$ with 11 values (for the low- and moderate-ionization absorbers) or 6 decades of $\xi$ with 20 values (for the high-ionization state absorber).   Thus, the sampling falls somewhat below our 5-samples-per-decade criterion and interpolation to intermediate ionization states may become inaccurate.   

Finally, we find the need to include a soft excess component --- following previous works, we employ a purely phenomenological approach.  Previous treatments of the soft excess in NGC~3783 have used a blackbody component (characterized by temperature $T_{\rm bb}$) or Comptonization by ``warm" plasma (described using the {\tt comptt} model in XSPECv12.7.0 and characterized by a temperature $T_{\rm compt}$ and an optical depth $\tau$).    To be conservative, we initially included both a blackbody and a warm thermal Comptonization component in our spectral fit.   In all cases presented in this paper, the Comptonization component would drop out of the fit (with its normalization becoming zero).   Experimenting with including either the blackbody {\it or} the Comptonization component  suggest that the blackbody fit was slightly preferred ($\Delta\chi^2\sim 3$); in both cases, the components have essentially the same shape and contribute only to the lowest energies in the {\it Suzaku} energy band.   Henceforth, all of the spectral modeling in this paper uses a blackbody model for the soft excess.   Using spot-checks, we have confirmed that the key results (e.g. the conclusions regarding black hole spin) are unchanged by switching to a Comptonization realization for the soft excess.  

Here, we shall employ three realizations of this spectral model that differ in their treatment of iron abundance.   In Model~A, we assume that the circumnuclear region (inner accretion disk, outer accretion disk, and the obscuring torus) are chemically homogeneous.   We make no assumption, however, about the value of this common iron abundance; we allow it to be a free parameter ($Z_{\rm Fe,disk}=Z_{\rm Fe,tor}={\rm free}$).    Model~B also assumes a chemically homogeneous system but imposes the additional constraint that the iron abundance is solar ($Z_{\rm Fe,disk}=Z_{\rm Fe,tor}=1$).    Model~C describes a chemically inhomogeneous system in which the distant reflector (the torus or the outer accretion disk) are forced to have solar iron abundance but the inner accretion disk has a free iron abundance ($Z_{\rm Fe,disk}={\rm free}$, $Z_{\rm Fe,tor}=1$).   

\section{Results}

\subsection{The need for super-solar iron abundances and rapid black hole spin}
\label{sec:bestfit}

\begin{figure*}
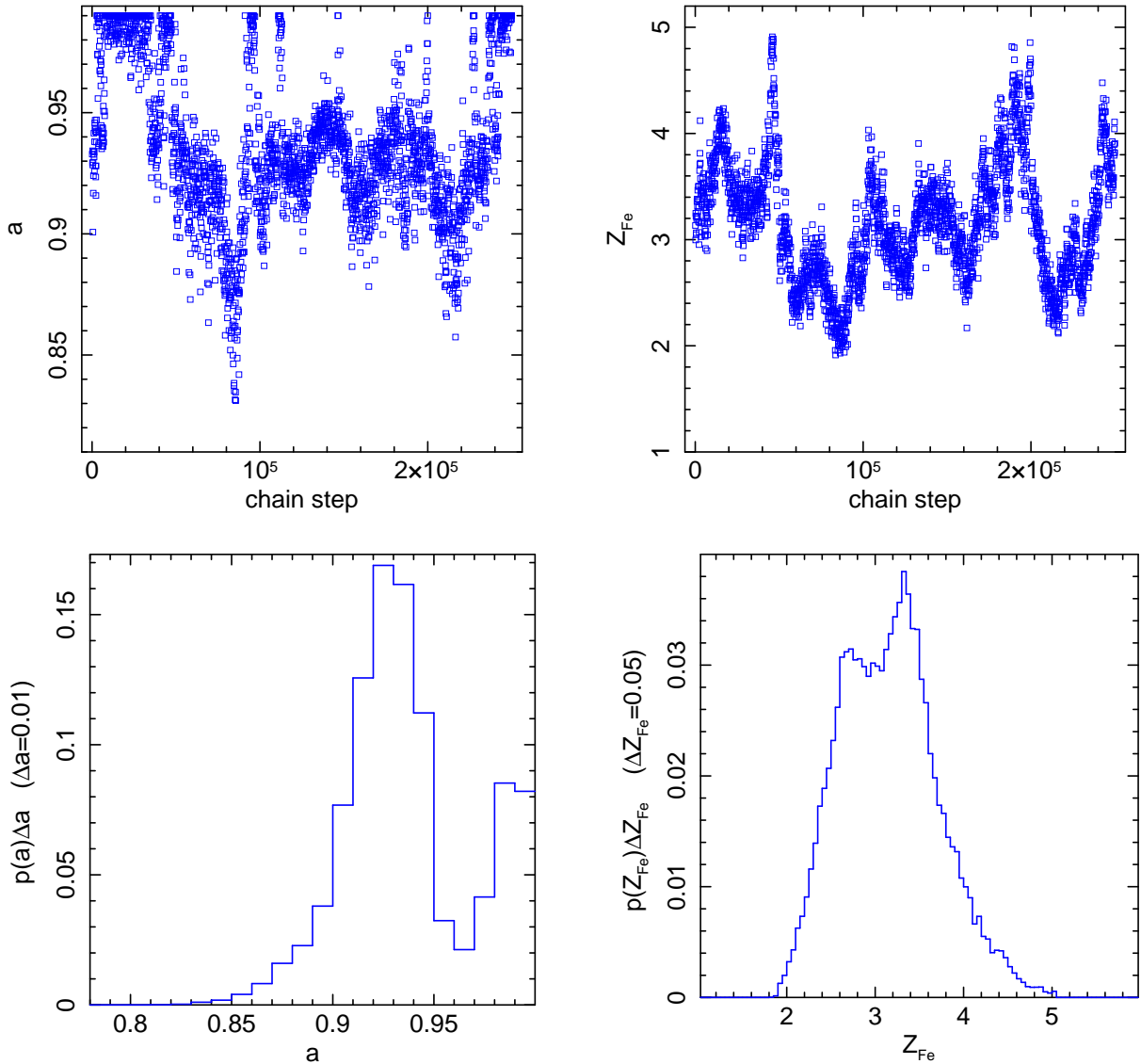

\hbox{
\psfig{figure=f2a.ps,width=0.4\textwidth,angle=270}
\hspace{1cm}
\psfig{figure=f2b.ps,width=0.4\textwidth,angle=270}
}
\vspace{0.5cm}
\hbox{
\psfig{figure=f2c.ps,width=0.4\textwidth,angle=270}
\hspace{1cm}
\psfig{figure=f2d.ps,width=0.4\textwidth,angle=270}
}
\caption{MCMC results for black hole spin and iron abundance from Model~A.  {\it Top panels : }Parameter value along the 250,000-element chain trajectory for the black hole spin $a$ (left) and iron abundance $Z_{\rm Fe}$ (right).   For clarity, we only show every 100th element of the chain.   {\it Bottom panels : }Probability distributions for the black hole spin and iron abundance.}
\label{fig:mcmc_best}
\end{figure*}

\begin{table*}[t]
\centerline{
\begin{tabular}{llccc}\hline\hline
Model Component & Parameter & Model A & Model B & Model C \\
& & (Homo; free) & (Homo; solar) & (Inhomo) \\\hline
Galactic column & $N_H$ & 9.91(f) & 9.91 (f) & 9.91 (f) \\\hline
Warm Absorbers & 	$N_1$ & 		$98^{+33}_{-16}$ 		& $91^{+22}_{-39}$			&	$101^{+16}_{-22}$	\\
			&	$\log\xi_1$ & 	$1.47\pm 0.05$ 		& $1.45^{+0.05}_{-0.10}$ 	&	$1.47\pm 0.05$	\\
&				$N_2$ & 		$192_{-72}^{+3}$ 		& $234^{+47}_{-36}$ 		&	$173^{+37}_{-20}$	\\
&				$\log\xi_2$ & 	$1.92_{-0.07}^{+0.03}$	& $1.93_{-0.01}^{+0.02}$		&	$1.93^{+0.02}_{-0.10}$	\\
&				$N_3$ & 		$217^{+65}_{-52}$		& $269^{+42}_{-52}$ 		&	$233^{+55}_{-63}$	\\
&				$\log\xi_3$ & 	$2.56^{+0.05}_{-0.04}$	& $2.60^{+0.03}_{-0.05}$ 	&	$2.58^{+0.05}_{-0.07}$	\\\hline
Power-law & 		$\Gamma$ & 	$1.79_{-0.03}^{+0.04}$  	& $1.85^{+0.03}_{-0.02}$ 	&	$1.80\pm 0.03$	\\
& 				$A_{\rm pl}$ & $1.42\pm 0.05$	& $1.62\pm 0.07$					&	$1.46\pm 0.05$	\\\hline
Blackbody & 		$kT_{\rm bb}$ & $0.054^{+0.002}_{-0.003}$ & $0.049^{+0.02}_{-0.03}$	&	$0.052\pm 0.003$	\\
& 				$A_{\rm bb}$ & $1.9^{+1.4}_{-0.5}$ 		& $4.4^{+5.3}_{-1.8}$	&	$2.4^{+1.8}_{-0.9}$	\\\hline
Neutral reflection & 	${\cal R}$ & 	$0.43^{+0.08}_{-0.06}$ 	& $0.83^{+0..07}_{-0.06}$ 	&	$0.78^{+0.08}_{-0.06}$	\\
(``torus") &		$Z_{\rm Fe,tor}$ & $3.3^{+0.8}_{-1.0}$ 	& 1 (f)					&	$1(f)$	\\\hline
Scattering fraction &$f$ & 		$0.20\pm 0.02$ 		& $0.20\pm 0.02$ 			&	$0.21\pm 0.02$	\\\hline
{\sc Fe\,xxvi} line (GAU) & $E$ & 	6.97(f) 				& 6.97(f)					&	6.97(f)	\\
& 				$\sigma$ &	 0.0154(f) 			&  0.0154(f) 				&	0.0154(f) 	\\
& 				$W_{Fe26}$ & $20\pm 5$ 				& $20\pm 5$						&	$20\pm 5$	\\\hline
Accretion disk & 	$Z_{\rm Fe,disk}$ & $=Z_{\rm Fe,tor}$ 	& 1 (f) 					&	$4.2^{+0.8}_{-1.3}$	\\
& 				$\xi$ & 		$<8$					& $<5$					&	$<10$	\\
& 				$i$ & 		$24^{+2}_{-3}$ 		& $22\pm 4$				&	$22^{+2}_{-3}$	\\
& 				$r_{\rm in}$ & 	$r_{\rm isco}$\,(f) 		& $r_{\rm isco}$\,(f) 			&	$r_{\rm isco}$\,(f)	\\
& 				$q_1$ & 		$6.6^{+0.9}_{-1.4}$ 		& $5.0^{+1.8}_{-3.8}$		&	$5.3^{+0.7}_{-1.9}$	\\
& 				$r_{\rm br}$ & 	$4.3^{+1.2}_{-0.6}r_g$ 	&  $1.5^{+3.8}_{-0.4}r_{\rm isco}$&	$5.2\pm 2.0r_g$	\\
& 				$q_2$ & 		$2.9\pm 0.2$			& $2.5^{+1.2}_{-0.9}$		&	$2.7\pm 0.3$	\\
& 				$r_{\rm out}$ & $400r_g$\,(f) 			& $400r_g$\,(f) 			&	$400r_g$\,(f)	\\\hline
SMBH spin & 		$a$ & 		$0.92^{+0.07}_{-0.03}$ 	& $0.59^{+0.38}_{-1.47}$ 	&	$0.90^{+0.09}_{-0.30}$\\\hline
$\chi^2/\nu$ & 		& 			$917/687$ 			& $953/688$ 				&	$931/687$	\\\hline
\end{tabular}
}
\caption{Spectral parameters for the model fits discussed in this paper.  Column densities are in units of $10^{20}\pcmsq$.   Ionization parameters are in cgs units.  The power-law normalization is in units of $10^{-2}\,{\rm ph}\,{\rm s}^{-1}\,{\rm keV}$ measured at 1\,keV.   The blackbody normalization is in units of $10^{-2}\,{\rm ph}\,{\rm s}^{-1}$, and its temperature is in units of keV.    The energy and 1-$\sigma$ width of the Fe26 line is in units of keV and the equivalent width $W_{\rm Fe26}$ is in units of eV.   Fixed parameters are denoted by ``(f)".  All errors are quoted at the 90\% confidence level.}
\label{tab:fit_results}
\end{table*}

Model~A is essentially identical to the spectral model of B11 and a traditional $\chi^2$ analysis of the {\it Suzaku} spectrum with this model was published in that paper.   As already noted in the introduction, B11 finds a rapidly spinning black hole and super-solar iron abundance.    However, a significant concern when $\chi^2$ fitting a spectral model with a large number of free parameters ($N=22$ free parameters in our case, including all model normalizations and the FI/XIS1 cross-normalization factor) is that we will mistake a local minimum in $\chi^2$ space for the global best fit.   Furthermore, the spectral model may genuinely possess multiple solutions and traditional $\chi^2$ error analysis is not well suited for mapping out multi-peaked probability distributions for model parameters.  

To address these limitations and give an independent validation of the statistical analysis of B11, we employ MCMC techniques which are now implemented in the {\sc Xspec} spectral fitting package.  Operationally, we proceed as follows.   Using traditional $\chi^2$  minimization, we obtain the best fit of Model~A with the full multi-instrument {\it Suzaku} dataset.  We then run five 55,000 element chains, starting from a random perturbation away from the best-fit and ignoring (``burning") the first 5000 elements of each chain.  The proposal for the chain (i.e. probability distribution that dictates the Monte Carlo step for each model parameter as the chain proceeds) is drawn from the diagonal of the covariance matrix resulting from the initial $\chi^2$ minimization, assuming Gaussian probability distributions with a rescaling factor of $10^{-3}$.   

Figure~\ref{fig:mcmc_best} (top panels) shows the values of the spin and iron abundance through the 250,000 elements of the chains Model A, showing every 100th element for clarity.  The lack of large excursions or trends suggests that we have indeed found the global best fit of this spectral model.  Experiments in which we start at a distant point in parameter space (e.g., setting initial seeds of $a=-0.95$ and $Z_{\rm Fe}=1$) converge on the same best fit.   Similarly, employing a different proposal (using Cauchy probability distribution or a larger rescaling factor) also converges on the same best fit, albeit slowly and with a very high repeatability fraction in the chain.    

The best-fit and MCMC-derived (90\% confidence level) errors are reported in Table~\ref{tab:fit_results}, and Fig.~\ref{fig:mcmc_best} (bottom panels) reports the detailed probability distributions for the black hole spin and iron abundance resulting from Model~A.   Results are, within the error bars, completely consistent with the results of B11\footnote{We note that goodness of fit for this model is formally unacceptable ($\chi^2/{\rm dof}=1.33$ for 687 degrees of freedom).  However, detailed visual inspection of the model fit to the folded spectrum shows that this is due to residuals in the 0.7--1.5\,keV range; removing these energies from consideration results in a formally acceptable fit $\chi^2/{\rm dof}=1.03$ (556 degrees of freedom).   The fact that the soft X-ray residual features are narrower in energy than the instrumental response, and are uncorrelated between XIS-FI and XIS1, strongly points to calibration features.  However, the small amplitude of these residuals ($<5$\%) and the fact that the model averages between the two XIS spectra gives us confidence that they are not dramatically skewing our best fit model.}.      From the relativistically smeared disk features, we derive a black hole spin of $a>0.89$, disk inclination between $i=21-26^\circ$, and an iron abundance between $Z=2.3-4.1Z_\odot$.  The inner emissivity index is steep ($q_1=6.6^{+0.9}_{-1.4}$), in agreement with B11 and consistent with a rapidly spinning black hole.   The emissivity profile flattens to $q_2=2.9\pm 0.2$ beyond the break radius ($r_{\rm br}=3.7-5.5r_g$).   Thus, without any prior constraint or bias, the outer emissivity index is consistent with the fiducial value of $q_2=3$ as predicted by either a lamp-post illumination model or any model in which the X-ray irradiation/emissivity tracks the underlying dissipation in the accretion disk.  

\begin{figure*}[t]
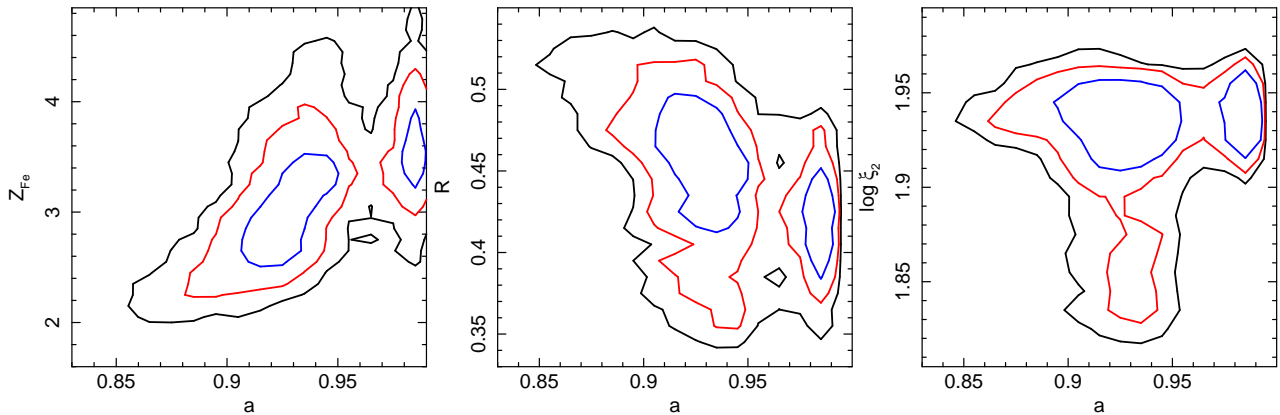

\hbox{
\psfig{figure=f3a.ps,width=0.3\textwidth,angle=270}
\psfig{figure=f3b.ps,width=0.3\textwidth,angle=270}
\psfig{figure=f3c.ps,width=0.3\textwidth,angle=270}
}
\caption{2-dimensional probability density functions of those parameters that show a correlation with spin.  {\it Left panel : }Probability density on the $(Z_{\rm Fe}, a)$-plane showing a positive correlation.   Contour levels are shown at $p(Z_{\rm Fe}, a)=1,3.3,10.0$ [defined such that the probability of being in the range $Z_{\rm Fe}\rightarrow Z_{\rm Fe}+\delta Z_{\rm Fe}$ and $a\rightarrow a+\delta a$ is $p(Z_{\rm Fe}, a)\delta Z_{\rm Fe}\delta a$].   {\it Middle panel : }Probability density on the $({\cal R}, a)$-plane revealing a weak negative correlation.   Contour levels are shown at $p({\cal R}, a)=10,33,100$. {\it Right panel : }Probability density on the $(\log\xi, a)$-plane revealing a weak negative correlation.   Contour levels are shown at $p(\log\xi, a)=10,33,100$.}
\label{fig:correlation}
\end{figure*}

An advantage of MCMC techniques is the efficiency with which statistical correlations between parameters can be uncovered.  Here, we are particularly interested in parameters that statistically correlate with black hole spin.  The strongest such correlation is between spin and iron abundance.   We illustrate this coupling in Fig.~\ref{fig:correlation} (left) which shows the 2-dimensional probability distribution on the $(Z_{\rm Fe}, a)$-plane for our Model~A analysis.   There is clearly a positive correlation between these two parameters, with more rapid spins simultaneously favoring higher iron abundance.   It is straightforward to understand this correlation.  As we discuss further in Section~\ref{sec:driver}, the sensitivity to spin comes as much from the need not to over-predict the blue horn of the iron line as it does from the need to capture the red-wing of the line.  As the iron abundance of the model increases, the ratio of iron line photons to reflection continuum increases.  In turn, the fit must blur the iron line more severely in order to not over-model the data in the 5--6\,keV band, and hence the inferred spin increases.    Similar reasoning explains the weak anti-correlation between the reflection fraction of the distant reflector ${\cal R}$ and the black hole spin (Fig.~\ref{fig:correlation}, middle).   As ${\cal R}$ decreases, the contribution of the inner accretion disk to the observed reflection spectrum increases leading to more iron line photons from the disk.  Consequently, the model needs to blur the disk line more severely (pushing to higher spin) in order to not over-model the data in the 5--6\,keV band.   

A less obvious statistical coupling is between the spin and and the ionization state of the intermediate warm absorber ($\xi_2$; Fig.~\ref{fig:correlation} right).   The 90\% confidence range of this parameter is $\log\xi_2=1.85-1.92$ (cgs units).  We find that the spread in allowable black hole spins is significantly greater when $\log\xi_2>1.9$, resulting in a ``T" shape to the 2-d probability on the $(\log\xi_2,a)$-plane.   No similar coupling is seen with $\xi_1$ or $\xi_3$.   We conclude that, in the range $\log\xi_2=1.9-2.0$, the spectral curvature can be (in part) degenerately described by either strong relativistic blurring characteristic of a very rapidly rotating black hole or by absorption edges from this the warm absorber (see Young et al. 2005 for a detailed discussion of these ``broad line mimicking warm absorbers").   The important thing to note is that, even accounting for all of these statistical correlations and degeneracies, the spectral fit still strongly requires a rapidly rotating black hole.

In addition to these statistical correlations, we note that both the 1-d and 2-d probability distributions (Figs.~\ref{fig:mcmc_best} and \ref{fig:correlation}) show double maxima, with peaks at $a\approx 0.93$ and $a=0.99$.   We have confirmed that the relativistic convolution model varies smoothly in the expected manner across this parameter range and, hence, is not an obvious source for this bi-modality.   We suspect that small ($\lesssim 1\%$) systematic errors in the effective area models (especially at energies characteristic of the iron line red-wing, 3--4\,keV) have induced this structure in probability space.   

\subsection{Models with imposed abundance constraints}

Having noted the existence of a correlation between iron abundance and spin, we now make a connection with the work of P11 and explore a model (Model~B) in which solar abundances are imposed.   We have repeated our MCMC analysis, now employing Model~B, again running five 55,000 element chains with a 5000 element burn period.   

\begin{figure}[t]
\centerline{
\psfig{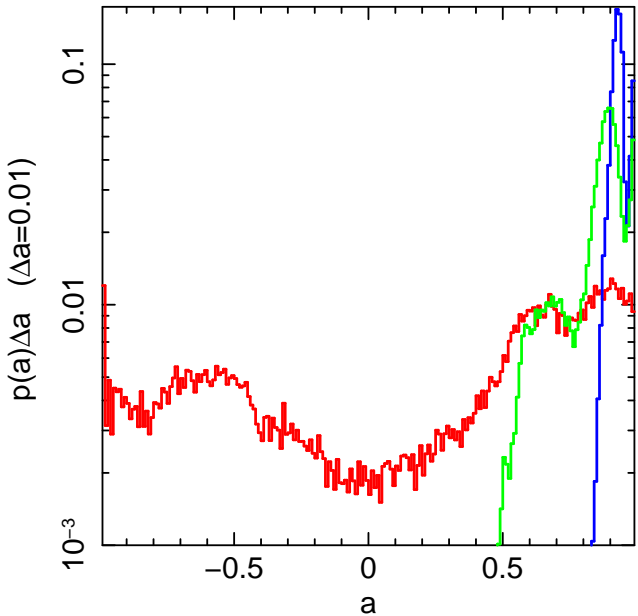}
}
\caption{Probability distribution for black hole spin in our fiducial model (Model A, blue), the solar abundance model (Model B, red), and the inhomogeneous-abundance model (Model C, green).}
\label{fig:comparison_spin}
\end{figure}

Imposing this abundance constraint (and refitting all other parameters) worsens the goodness of fit parameter by $\Delta\chi^2=36$ (for one fewer degree of freedom) and significantly changes the probability distribution for several of the key spectral parameters (Table~\ref{tab:fit_results}).  The principle effect is that, due to the decreased iron abundance, the fit demands a significantly higher neutral reflection fraction ${\cal R}$ in order to fit the narrow iron line.   This results in a slightly steeper best-fitting power-law continuum ($\Gamma$ steepens from 1.79 to 1.85).  At the same time, the ratio of iron line photons to reflection continuum from the accretion disk is also decreased by the reduced abundance.   Thus, the need to severely broaden the iron line and, hence, invoke a rapidly spinning black hole, is diminished.    In detail, the probability distribution for spin in this constrained fit is weakly bi-modal, with modest-to-rapid prograde {\it or} retrograde black holes being preferred over slowly spinning holes (Fig.~\ref{fig:comparison_spin}).

Both Model~A and Model~B assume that the distant reflector and the inner accretion disk have the same iron abundance.  In our final spectral model, Model~C, we explore the effects of imposing a mixed abundance model such that the distant reflector has a solar iron abundance but we allow the inner accretion disk to have a freely-fitting iron abundance.  In terms of goodness of fit, this is intermediate between Model~A and Model~B, with $\Delta\chi^2=14$ compared with our fiducial fit.  The MCMC analysis (executed in an identical manner to that for Model~A) shows that a rapidly spinning black hole is preferred, but there is a low-spin tail to the probability distribution giving a final result of $a>0.60$ (at 90\% CL).

\subsection{What drives the spin constraint?}
\label{sec:driver}

\begin{figure*}[t]
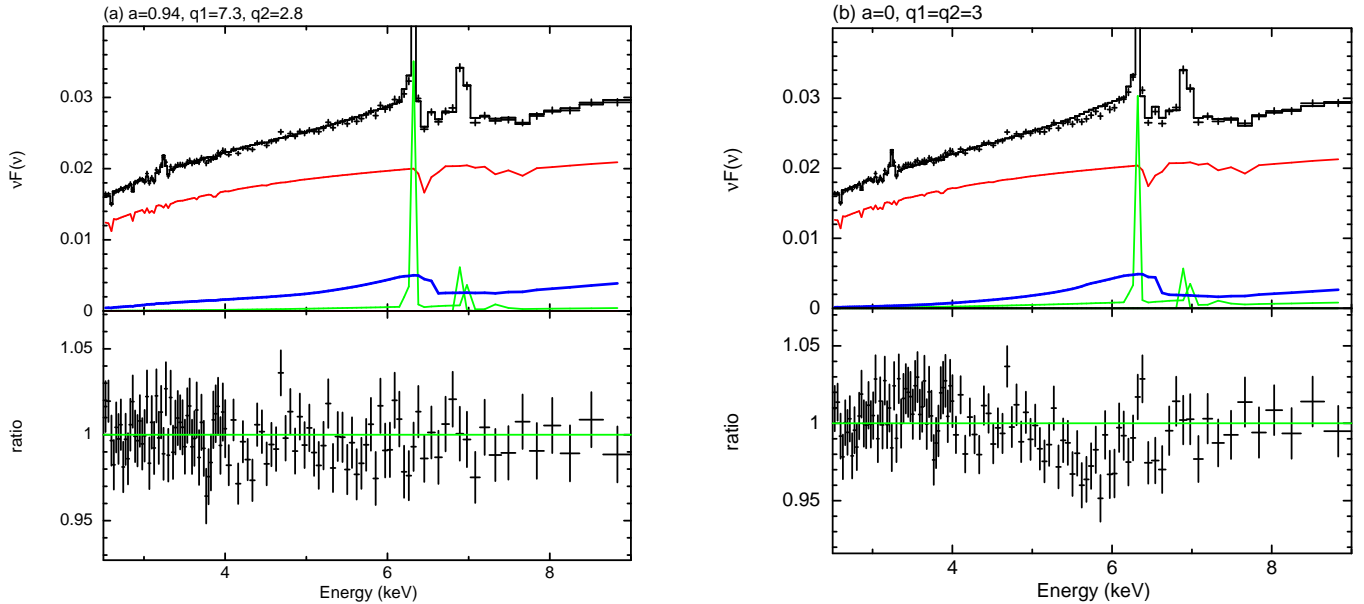

\hbox{
\psfig{figure=f5a.ps,width=0.45\textwidth,angle=270}
\hspace{1cm}
\psfig{figure=f5b.ps,width=0.45\textwidth,angle=270}
}
\caption{{\it Left panel : }Unfolded XIS data, best fitting model and resulting featureless residuals for Model~A.  The model components visible here are the absorbed power-law continuum (red), distant reflection (green) and relativistically smeared disk reflection (blue).   {\it Right panel : } Unfolded XIS data, best fitting model and residuals when the spin is constrained to be $a=0$ and the emissivity indices are constrained to be $q_1=q2=3$ (but all other parameters are refitted).   Model components are colored the same as above.}
\label{fig:spin_residuals}
\end{figure*}

Fundamentally, analyses of disk reflection features diagnose black hole spin using the fact that the observed broadening/redshifting can be so extreme as to demand reflection from within $6r_g$, the ISCO of a non-rotating black hole (Reynolds \& Fabian 2008).  Thus, one might expect that the spin constraints come principally from the modeling of the extreme red-wing of the iron line and that, if the spectral model being compared with the data has too low of a spin, the main result would be positive residuals (model under-predicting data) at energies characteristic of the red-wing (3--5\,keV).   In fact the situation is more subtle due to the ability of other model components and parameters (e.g. the warm absorber, or the iron abundance of the accretion disk) to compensate for an incorrect spin.

We illustrate this in Fig.~\ref{fig:spin_residuals}.   Figure~\ref{fig:spin_residuals}a shows the model and residuals for the best fit model presented in Section~\ref{sec:bestfit}.   For comparison, Fig.~\ref{fig:spin_residuals}b shows the model and residuals for a fit in which the spin has been fixed to $a=0$ and the emissivity indices have been fixed to $q_1=q_2=3$ (to be physically consistent with a non-spinning black hole).  While there are weak positive residuals in extreme wing of the iron line at 3--4\,keV, the most significant deviations are {\it negative} residuals at 5--6\,keV.  This can be understood as follows.  The overall strength of the accretion disk reflection as well as the iron abundance of the disk is, to a major extent, determined by the iron-edge (rest frame energy 7.1\,keV) and the form of the hard band (7--40\,keV) spectrum.  The question then becomes whether an iron line with the self-consistently determined equivalent width is compatible with the data.  In the case of NGC3783, we find the resulting iron line is too strong (over-predicts the data) if it is subjected to blurring from around a non-spinning black hole.  Only the enhanced relativistic effects of a rapidly-rotating black hole can distribute the line flux sufficiently to make the iron line compatible with the data.

Figure~\ref{fig:spin_residuals}b teaches us an important lesson about the role of possible systematic errors.   Specifically, given data of realistic complexity, the {\it relative} effective area calibration of a medium-resolution instrument such as the XIS must be accurate to better than 2\% across the 3--7\,keV band if we hope to detect the signatures of black hole spin.  A more complete study of the role of systematic errors in black hole spin measurements (including an assessment of how the effective area calibration requirements relax when one has simultaneous high-resolution constraints on the warm absorber, as will be the case for the Soft X-ray Spectrometer [SXS] on {\it Astro-H}) will be presented in a future publication (A.Lohfink in prep.).

\section{Discussion}

\subsection{Summary of observational results}

We have revisited the long 2009 {\it Suzaku} data for the Seyfert 1.5 galaxy NGC~3783 in the light of conflicting published results for the black hole spin.   We find that the treatment of the iron abundances in both the inner accretion disk and the distant torus are important for the derived SMBH spin.   We expect that the timescale on which stellar processes enrich their environment ($\sim 10^8$\,yr) is longer than the timescale on which material is exchanged between the torus and the inner disk (of order the dynamical timescale of the torus, $10^5$\,yr) leading to the conclusion that the central engine will be chemically homogeneous.  Employing this homogeneity assumption ($Z_{\rm Fe,disk}=Z_{\rm Fe,tor}$), our spectral analysis shows that the data strongly require high iron abundance $Z_{\rm Fe}=2-4Z_\odot$ and a rapidly spinning black hole $a>0.88$.  We note that such a metallicity enhancement is completely in line with that inferred from optical/UV studies of the broad line region (BLR) of type-1 AGN, e.g. see Warner et al. (2004) and Nagao, Marconi \& Maiolino (2006).     In agreement with previous analyses, we show that constraining the abundance to be solar worsens the goodness of fit and leaves the SMBH spin unconstrained.    Employing a chemically-inhomogeneous model ($Z_{\rm Fe,tor}=Z_\odot$; $Z_{\rm Fe,disk}=${free}) we show that it is the abundance of the accretion disk component that is primarily (but not exclusively) driving the spin constraint.   

\subsection{Iron abundance measurements from X-ray reflection}

Even if the accretion disk and the surrounding circumnuclear region are chemically-homogeneous, there are subtleties that can affect the abundances that are {\it inferred} from X-ray reflection fits.  Firstly, let's consider the ``cold" reflection from the torus.  Fundamentally, X-ray reflection models constrain iron abundance by comparing the equivalent width of the fluorescent iron line to the strength of the Compton reflection continuum at $E>7\keV$.   With photons at these energies, the nature of the Compton scattering and hence the reflection continuum is the same whether the bulk of the electrons are free or bound into hydrogen/helium atoms.   However, of potential importance is the depletion of iron into dust grains within the torus.  While an iron atom can readily undergo K-shell fluorescence even when bound into a solid, sufficiently large dust grains will lead to the self-shielding of iron and hence a reduction in the iron-K$\alpha$ equivalent width.   The minimum grain size $x_{\rm min}$ for which K-shell self-shielding is important is given by assuming pure iron, $x_{\rm min}=m_{\rm Fe}/\rho_{\rm Fe}\sigma_{\rm th}$ where $m_{\rm Fe}$ is the mass of a iron atom, $\rho_{\rm Fe}$ is the density of the grain, and $\sigma_{\rm th}$ is the cross-section at the K-shell photoionization threshold for cold iron.  Evaluating, using $\rho_{\rm Fe}=7.7\times 10^3\,{\rm kg}\,{\rm m}^{-3}$ and $\sigma_{\rm th}=3.5\times 10^{-24}\,{\rm m}^{-2}$, we obtain $x_{\rm min}\approx 3.5\,\mu{\rm m}$.   The minimum grain size may be up to twice this size is the iron if bound into a compound (e.g. an oxide, silicate or sulphide), and even more if the grain has an amorphous structure.   Thus, self-shielding can only reduce the iron-K$\alpha$ equivalent width if the dust mass is dominated by very large grains.

For the innermost accretion disk, different considerations are at play.  Since there are clearly no issues of dust depletion in this hot plasma, the normal assumption is that the abundances measured for the inner accretion disk should faithfully represent the true abundances.   However, in this extreme environment, there are some interesting mechanisms that have the potential to affect the {\it photospheric} iron abundance of the accretion disk.  Skibo (1997; also see Turner \& Miller 2010) suggests that the bombardment of the inner accretion disk by energetic (relativistic) protons can lead to the spallation of iron, enhancing the abundance of Ti, V, Cr and Mn at the expense of the iron abundance.  However, significant iron spallation within the disk only occurs if a large fraction of the accretion energy ($\sim 0.1\dot{M}c^2$) is carried by relativistic protons which subsequently bombard the disk.   The resulting $\gamma$-ray emission (resulting from the decay of pions produced by these interactions) was estimated by Skibo (1997) and violates the $4\sigma$ upper limits for radio-quiet Seyfert galaxies recently found Teng et al. (2011) using {\it Fermi} data.  

Most AGN in which X-ray reflection from the inner accretion disk has been modeled in detail have been found to possess supersolar iron abundance (e.g., see Table~2 in B11).  In most cases, such as NGC~3783, the inferred abundances are completely consistent with that found for the BLR and hence probably reflects a true-chemically enriched circumnuclear environment.   In a small number of cases, however, rather extreme disk iron abundances are inferred (e.g., 1H0707$-$495 in which a strong iron-L line implies iron abundances of $>7Z_\odot$; Zoghbi et al. 2010).   In the rest of this Section, we discuss a mechanism, radiative levitation, that has the potential to significantly enhance the {\it photospheric} iron abundance of the inner disk.

\subsection{Radiative levitation in AGN accretion disks}\label{sec:radlev}

In moderate luminosity AGN such as NGC~3783, the inner regions of the accretion disk are radiation-pressure dominated with a photospheric temperature of $T\sim 10^5-10^6$\,K (Shakura \& Sunyaev 1973).  By definition, this implies that each radius of the disk is at the ``local" Eddington limit in the sense that the radiation pressure force acting on the plasma is balancing the vertical component of gravity.  However, at the microphysical level, the radiative force on metal ions can be very different to the radiative force on the background, fully ionized, hydrogen/helium plasma.    In particular, since iron can possess moderate-to-low charge states ({\sc Fe\,xvii} and below) with populated L- and M-shells, the radiation force of the $10^5-10^6$\,K thermal radiation pushing on the UV/EUV iron absorption lines and the EUV/X-ray bound-free edges can produce a net upwards force on the iron ions many orders of magnitude greater than the vertical gravity.   In the absence of other dynamics (but see below), this radiative pushing will cause the iron to diffuse upwards towards the photosphere and could significantly enhance the iron abundance of the disk surface.   This phenomenon  has been discussed in the context of the surface abundances of hot white dwarfs (Chayer, Fontaine \& Wesemael 1995; Seaton 1996, Wassermann et al. 2010), and opacity enhancement from radiative levitation have been identified as key to understanding the 1--2\,min oscillations in the unusual sub-dwarf O-star SDSS J160043.6+074802.9 (Fontaine et al. 2008).  However, to the best of our knowledge, radiative levitation has not been discussed previously in the context of AGN accretion disks.   Here, we conduct a preliminary assessment of whether radiative levitation can be relevant for AGN accretion disks --- we conclude that it may be relevant, and lay out a path for a more rigorous treatment of its effect.

Consider an iron ion (mass $m_{\rm Fe}$ and charge state $Q_{\rm Fe}$) that is within a few Thomson depths of the photosphere of the inner accretion disk at radius $R=xr_g$ from the black hole.  The dimensionless accretion rate of the accretion disk is taken to be $\dot{m}\equiv L/\eta L_{\rm Edd}$ where $L_{\rm Edd}$ is the Eddington luminosity and $\eta$ is the accretion efficiency.  The standard theory of radiation-dominated disks (Shakura \& Sunyaev 1973) then gives the disk thickness as $h={3\over 2}\dot{m}r_g$.   Our iron ion is subject to a gravitational force pulling it towards the midplane of the disk, $F_{\rm grav}=m_{\rm Fe}g$ where $g=GMh/R^3$.   The ion is also subject to a radiative force pushing it upwards towards the photosphere $F_{\rm rad}\equiv m_{\rm Fe}g_{\rm rad}$, as well as the effect of Coulomb collisions with the ions in the background plasma (assumed, for simplicity, to be pure hydrogen).  Following Seaton (1996), the (upward) drift velocity of the iron ion relative to the background plasma is given by
\begin{equation}
v={\cal D}\left[\frac{m_{\rm Fe}}{kT}(g_{\rm rad}-g)-\frac{1}{\chi}\frac{\partial \chi}{\partial z}\right],
\end{equation}
where $T$ is the temperature of the plasma (assumed to be in LTE), $\chi$ is the abundance enhancement factor of the iron ions, and ${\cal D}$ the diffusion coefficient.  For this non-degenerate plasma, the appropriate form of the diffusion coefficient is (Chapman \& Cowling 1962; Alcock \& Illarionov 1980),
\begin{equation}
{\cal D}=\frac{3(2kT)^{5/2}}{16n(\pi m_p)^{1/2}Q_{\rm Fe}^2e^4\Lambda},
\end{equation}
where $n$ is the proton number density of the background hydrogen plasma and $\Lambda$ is the Coulomb logarithm given by,
\begin{equation}
\Lambda=\ln\left(1+\frac{(kT)^3}{4\pi nQ_{\rm Fe}^2e^6}\right).
\end{equation}

To assess the potential relevance of radiative levitation, we will estimate the drift timescale across a Thomson depth, $t_{\rm drift}=1/n\sigma_Tv$.   Competing mechanisms (e.g. ``stirring" of the surface layers) will occur on the order of the dynamical timescale $t_{\rm dyn}=\sqrt{R^3/GM}$, and thus we shall examine the ratio of timescales $t_{\rm drift}/t_{\rm dyn}$.     We will also limit our consideration to the situation where $\partial\chi/\partial z=0$ (i.e. no enhancement has yet occurred).   Using the above expressions and definitions, a little algebra shows that
\begin{equation}
\frac{t_{\rm drift}}{t_{\rm dyn}}=\frac{16}{9}\frac{(\pi m_p)^{1/2}e^4\Lambda}{m_{\rm Fe}\sigma_Tc}\,\frac{Q_{\rm Fe}^2}{(2kT)^{3/2}}\,\frac{x^{3/2}}{\dot{m}\Xi_{\rm rad}},
\end{equation}
where we have defined,
\begin{equation}
\Xi_{\rm rad}=\frac{g_{\rm rad}}{g}-1.
\end{equation}
Note that, because we are considering the drift timescale across a Thomson depth, the proton density of the plasma $n$ only appears in the expression for $t_{\rm drift}/t_{\rm dyn}$ via the Coulomb logarithm $\Lambda$; hence the result is extremely weakly dependent on the density.    Now, suppose that the net upward energy flux of radiation (as a function of frequency $\nu$) is $F_\nu$.  The net radiation force on the iron ion is
\begin{equation}\label{eqn:mfegrad}
m_{\rm Fe}g_{\rm rad}=\frac{1}{c}\int_0^\infty\sigma_\nu F_\nu\,d\nu,
\end{equation}
where $\sigma_\nu$ is the absorption cross-section of the ion.  Assuming that the opacity of the disk is dominated by electron scattering, we can use the fact that the disk is radiation pressure supported to write,
\begin{equation}
m_pg=\frac{\sigma_T}{c}\int_0^\infty F_\nu\,d\nu,
\end{equation}
which, when combined with eqn.~\ref{eqn:mfegrad} gives,
\begin{equation}
\Xi_{\rm rad}=\frac{m_p}{m_{\rm Fe}}\frac{{1\over\sigma_T}\int_0^\infty\sigma_\nu F_\nu\,d\nu,}{\int_0^\infty F_\nu\,d\nu}.
\end{equation}

\begin{figure}[t]
\centerline{
\psfig{figure=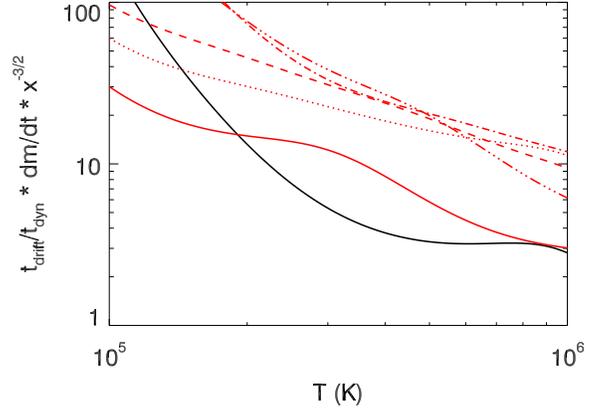,width=0.5\textwidth}
}
\caption{Estimates of the radiative levitation drift timescale for various charge states of iron (expressed as the ratio $t_{\rm drift}/t_{\rm dyn}\times \dot{m}x^{-3/2}$), calculated under the assumptions laid out in \S\ref{sec:radlev}.   Shown here is {\sc Fe\,viii} (black; solid), {\sc Fe\,xvii} (red; solid), {\sc Fe\,xviii} (red; dotted), {\sc Fe\,xix} (red; dashed), {\sc Fe\,xx} (red; dot-dashed), and {\sc Fe\,xxi} (red; dot-dot-dot-dashed).}
\label{fig:rad_lev}
\end{figure}

A full evaluation of $F_\nu$ and hence $\Xi_{\rm rad}$ must be performed in the context of a detailed radiative transfer model; such a treatment is beyond the scope of this paper.   Here, we make an enormous simplifying assumption, that the spectral shape of $F_\nu$ is proportional to the Planck function $B_\nu(T)$.  While this is certainly inappropriate, it does allow us to study whether radiative levitation has the {\it potential} to be important.   We also assume that only opacities from bound-bound transitions are relevant.   Under these assumptions, a little algebra shows that
\begin{equation}
\Xi_{\rm rad}=\frac{45}{8\pi^4}\frac{m_em_pc^3h}{m_{\rm Fe}e^2kT}\left(\sum_if_i\frac{y_i^3}{e^{y_i}-1}\right)-1,\label{eqn:rad_force}
\end{equation}
where we have replaced the integral in eqn.~\ref{eqn:mfegrad} with the sum over bound-bound transitions with oscillator strengths $f_i$ and normalized energies $y_i\equiv h\nu_i/kT$.   Given the appropriate atomic data, we now have all of the ingredients required to compute $t_{\rm drift}/t_{\rm dyn}$ as a function of charge-state $Q_{\rm Fe}$, temperature $T$, dimensionless accretion rate $\dot{m}$ and dimensionless radius $x$.

Figure~\ref{fig:rad_lev} shows this ratio $t_{\rm drift}/t_{\rm dyn}$ (normalized to $x=\dot{m}=1$) as a function of temperature for various charge states of iron.   Specifically, we show the results for {\sc Fe\,viii} (the dominant charge state of iron for $\log\xi=0.4-1.2$) and {\sc Fe\,xvii}--{\sc Fe\,xxi} (the dominant charge states for $\log\xi=1.8-2.8$).   The atomic data needed to evaluate eqn.~\ref{eqn:rad_force} is obtained from the the XSTAR database.  It can be seen that, for $T>3\times 10^5$\,K, the drift timescale for {\sc Fe\,viii} and {\sc Fe\,xvii} can be only a factor of a few greater than the dynamical timescale.   For {\sc Fe\,xviii}--{\sc Fe\,xxi} ions, the drift timescale is approximately an order of magnitude greater than the dynamical timescale for $T=10^6$\,K, but becomes rather longer for smaller temperatures.   We conclude that radiative levitation {\it may} be an important process in the inner disks ($x\sim 1-{\rm few}$) of high accretion rate ($\dot{m}\sim 1$) low black hole mass (high disk temperature) AGN provided that the ionization state promotes certain charge states of iron.

The above calculation is only a plausibility argument.  Firstly, as mentioned already, a real radiative transfer calculation is required to compute the radiative force on the iron ions.   Line saturation will be extremely important and tend to decrease the upward radiative force, but this may be partially counteracted by ionization gradients and Doppler shifts associated with the radiation induced ion drift.  We also note that our calculation does not include the radiation force resulting from pushing on bound-free transitions for which saturation will be much less important.   Given sufficient time (many drift timescales), radiative levitation would achieve an equilibrium in which the iron enhancement produces increased opacity (self-shielding) such as to bring the radiative and gravitational forces into balance (Chayer, Fontane \& Wesemael 1995).   But, as a second major issue, we must characterize processes that will mix the plasma and erase the radiation-induced enhancements.   If the surface layers are turbulent with eddy timescales comparable to or less than the local dynamical timescale, we see that radiative levitation may be of limited importance.   On the other hand, if the surface layers of the disk are magnetically dominated, turbulence can be suppressed and there may be sufficient time for radiative levitation to operate.  Future radiation-MHD simulations of accretion disks with AGN-like parameters are required to assess whether the surface layers are likely to be well mixed.

\section{Conclusions}

Given its astrophysical importance, the observational determination of SMBH spin is clearly of great interest.  The study of X-ray reflection features from the innermost regions of SMBH accretion disks is, at the current time, the most promising technique for measuring SMBH spin and is a prime focus of future X-ray observatories such as {\it Astro-H} and {\it LOFT}.  Thus, now is the time to test and refine our methodologies for extracting robust and reliable black hole spins from the X-ray data.

The discrepant published values for the SMBH spin in NGC~3783 provide an ideal opportunity to examine and improve our methodology.  Here, we identify several factors that have led to this discrepancy.   For AGN that are strongly affected by a warm absorber, computational practicalities dictate that we describe the (possibly multiple) warm absorber zone(s) using tabulated grids of pre-calculated photoionization models.   We point out the importance of using grids that adequately sample parameter space, especially in the ionization parameter $\xi$ dimension.   We propose a 5-sample-per-decade criterion and show that the linear interpolation of warm absorber grids have significant inaccuracies if sampled at a significantly lower density.   Most importantly, however, we identify a statistical coupling between the iron abundance of the accretion disk and the inferred black hole spin.   We show that one must allow for the possibility of non-solar abundances.  Putting everything together, we find that the data for NGC3783 require a moderately supersolar iron abundance ($Z\approx 2-4Z_\odot$) and a rapidly rotating black hole ($a>0.89$).   

While the abundance enhancement required for NGC~3783 is modest and within the realm of metallicity estimates from BLR studies, we note that the X-ray analysis of some other AGN have found extreme iron abundances.  This prompts us to consider mechanisms that can enhance the {\it observed} abundances above and beyond the true macroscale abundances of the circumnuclear region.   We give a plausibility argument to show that radiative levitation, the radiation-driven diffusion of iron towards the disk photosphere, may operate in the innermost regions of high accretion rate AGN disks.    More theoretical work is required to assess whether radiative transfer effects and/or turbulent mixing of the photospheric regions of the disk will render radiative levitation ineffective.  However, even the current calculation makes predictions that can guide future observational study.  The clearest prediction is that there should be a positive correlation between the Eddington ratio of an AGN and the iron abundance inferred from X-ray reflection spectroscopy of the innermost disk.   This may show up in detailed studies of disk reflection across a sample of AGN, or detailed studies of a single source that is undergoing accretion rate changes.   Indeed, if we find an individual source that shows clear and compelling evidence for a change in its inner disk iron abundance, we will be forced into accepting that some mechanism (such as radiative levitation) is modulating the observed iron abundance away from that of the general environment.  

It is also interesting to consider the wider implications for extreme iron enhancements due to radiative levitation.   Since this mechanism only operates in the very central regions of the accretion disk ($t_{\rm drift}/t_{\rm dyn}\propto r^{3/2}$), it is likely to be ineffective at modifying the abundance of accretion disk winds that originate from larger radii (which manifest themselves as X-ray or UV absorbers).   However, the surface regions of the innermost disk are very likely the source of the ``funnel flow" which bounds the base regions of Blandford-Znajek powered relativistic jets (Blandford \& Znajek 1977; McKinney \& Narayan 2007).   Thus, especially in their base regions, Blandford-Znajek jets may entrain plasma that has been highly enriched in iron with potentially important consequences for high-energy hadronic processes within jets as well as the composition of high-energy cosmic rays that may originate from AGN jets.



\acknowledgments
\section*{Acknowledgments}
We thank Cole Miller, James Reeves and Adam Patrick for useful discussions during the course of this work.   We also thank the anonymous referee for their positive and insightful comments which significantly improved the paper.   This research has made use of data obtained from the High Energy Astrophysics Science Archive Research Center (HEASARC) provided by NASA's Goddard Space Flight Center.  CSR and AML thank support from the NASA Suzaku Guest Observer Program under grant NNX10AR31G.   MLT thanks support from NASA Long Term Space Astrophysics grant NAG513065.


\begin{thebibliography}{}

\bibitem[Alcock and Illarionov(1980)]{alcock80}
Alcock, C., Illarionov, A., 1980, ApJ, 235, 534

\bibitem[Ballantyne et al.(2002)]{ballantyne02}
Ballantyne, D. R., Ross, R. R., Fabian, A. C., 2002, MNRAS, 332, L45

\bibitem[Ballantyne et al.(2011)]{ballantyne11}
Ballantyne, D. R., McDuffie, J. R., Rusin, J. S., 2011, ApJ, 734, 112

\bibitem[Blandford and Znajek(1977)]{bz77}
Blandford R. D., Znajek R. L., 1977, MNRAS, 179, 433

\bibitem[Brenneman and Reynolds(2006)]{brenneman06}
Brenneman L.W., Reynolds C.S., 2006, ApJ, 652, 1028

\bibitem[Brenneman et al.(2011)]{brenneman11}
Brenneman L.W. et al., 2011, ApJ, 736, 103 (B11)

\bibitem[chapman and cowling(1952)]{chapman52}
Chapman S., Cowling T.G., 1952, Mathematical Theory of Non-Uniform Gases (Cambridge: Cambridge University Press)

\bibitem[Chayer et al.(1995)]{chayer95}
Chayer, P., Fontaine, G., Wesemael, F., 1995, ApJS, 99, 189

\bibitem[Dabrowski et al.(1997)]{dabrowski97}
Dabrowski Y. et al., 1997, MNRAS, 288, L11

\bibitem[Dauser et al.(2010)]{dauser10}
Dauser T., Wilms J., Reynolds C.S., Brenneman L.W., 2010, MNRAS, 409, 1534

\bibitem[Fabian et al.(1995)]{fabian95}
Fabian A.C. et al., 1995, MNRAS, 277, L11

\bibitem[Fontaine et al.(2008)]{fontaine08}
Fontaine, G., Brassard, P., Green, E. M., Chayer, P., Charpinet, S., Andersen, M., Portouw, J., 2008, A\&A, 486, L39

\bibitem[George and Fabian(1991)]{george91}
George I. M., Fabian A. C., 1991, MNRAS, 249, 352

\bibitem[Kallman and Bautista(2001)]{kallman01}
Kallman, T., Bautista, M. 2001, ApJS, 133, 221

\bibitem[Kaspi et al.(2002)]{kaspi02}
Kaspi, S., et al. 2002, ApJ, 574, 643

\bibitem[Komatsu et al. (2011)]{komatsu11}
Komatsu E., et al., 2011, ApJS, 192, 18

\bibitem[McKinney and Narayan(2007)]{mckinney2007}
McKinney J.C.,  Narayan R., 2007, MNRAS, 375, 513

\bibitem[Miller(2007)]{miller07}
Miller J.M., 2007, ARA\&A, 45, 441

\bibitem[Nandra et al.(2007)]{nandra07}
Nandra, K., OÕNeill, P. M., George, I. M., Reeves, J. N. 2007, MNRAS, 382,
194

\bibitem[nagao et al.(2006)]{nagao06}
Nagao, T.,Maiolino, R., Marconi, A., 2006, A\&A, 459, 85

\bibitem[Netzer et al.(2003)]{netzer2003}
Netzer H. et al., 2003, ApJ, 599, 933

\bibitem[Patrick et al.(2011)]{patrick11}
Patrick A.R. et al., 2011, MNRAS, 416, 2725 (P11)

\bibitem[Reis et al.(2012)]{reis12}
Reis R.C. et al., 2012, ApJ, 745, 93

\bibitem[Reynolds et al.(2012)]{reynolds12}
Reynolds C.S. et al., 2012, in proceedings of ``SUZAKU 2011: Exploring the X-ray Universe: Suzaku and Beyond", AIP Conference Proceedings, Volume 1427

\bibitem[Reynolds and Nowak(2003)]{reynolds03}
Reynolds C.S., Nowak M.A.., 2003, Phys. Rep., 377, 389

\bibitem[Reynolds and Fabian(2008)]{reynolds08}
Reynolds C.S., Fabian A.C., 2008, ApJ, 675, 1048

\bibitem[Reynolds et al.(1995)]{reynolds95}
Reynolds C. S., Fabian, A. C., Inoue, H., 1995, MNRAS, 276, 1311

\bibitem[Ross and Fabian(2005)]{ross05}
Ross R.R., Fabian A.C., 2005, MNRAS, 358, 211

\bibitem[Seaton(1996)]{seaton96}
Seaton, M. J., 1996, Ap\&SS, 237, 107

\bibitem[Skibo(1997)]{skibo97}
Skibo, J. G. 1997, ApJ, 478, 522

\bibitem[Tanaka et al.(1995)]{tanaka95}
Tanaka Y. et al., 1995, Nature, 375, 659

\bibitem[Teng et al.(2011)]{teng11}
Teng S., Mushotzky R.F., Sambruna R.M., Davis D.S., Reynolds C.S., 2011, ApJ, 742, 66

\bibitem[Theureau et al.(1998)]{theureau98}
Theureau, G., Bottinelli, L., Coudreau-Durand, N., Gouguenheim, L.,
Hallet, N., Loulergue, M., Paturel, G., Teerikorpi, P. 1998, A\&AS, 130,
333

\bibitem[Turner and Miller(2010)]{turner10}
Turner, T. J., Miller, L., 2010, ApJ, 709, 1230

\bibitem[warner et al.(2004)]{warner04}
Warner C., Hamann F., Dietrich M., 2004, ApJ, 608, 136

\bibitem[Wassermann et al.(2010)]{wassermann10}
Wassermann, D., Werner, K., Rauch, T., Kruk, J. W., 2010, A\&A, 524, 9

\bibitem[Zoghbi et al.(2010)]{zoghbi10}
Zoghbi A. et al., 2010, MNRAS, 401, 2419

\end{thebibliography}
\clearpage

\end{document}